\documentclass[english,jmp, reprint, amsmath, amssymb,reprint,]{revtex4-1}
\usepackage[T1]{fontenc}
\usepackage[latin9]{inputenc}
\usepackage{amssymb}
\usepackage{graphicx}
\usepackage{babel}
\usepackage{color}
\begin{document}

\title{Coulomb drag between carbon nanotube and graphene}

\author{Jean-Damien Pillet$^{1,2}$, Austin Cheng$^{3}$, Takashi Taniguchi$^{4}$,
Kenji Watanabe$^{4}$, Philip Kim$^{3}$}

\affiliation{$^{1}$Department of Physics, Columbia University, New York, New
York 10027, USA.}

\affiliation{$^{2}$Laboratoire des Solides Irradi\'{e}s, \'{E}cole Polytechnique, CNRS, CEA, Universit\'{e} Paris-Saclay, 91128  Palaiseau, France.}

\affiliation{$^{3}$Department of Physics, Harvard University, Cambridge, Massachusetts
02138, USA.}

\affiliation{$^{4}$National Institute for Materials Science, Namiki 1-1, Ibaraki
305-0044, Japan.}

\begin{abstract}
We report the observation of Coulomb drag between a two-dimensional (2D) electron gas in graphene and a one-dimensional (1D) wire composed of a carbon nanotube. We find that drag occurs when the bulk of graphene is conducting, but is strongly suppressed in the quantum Hall regime when magnetic field confines conducting electrons to the edges of graphene and far from the nanotube.   
Out-of-equilibrium and non-linear drag measurements show intriguing interplay between 1D and 2D conductors. These hybrid electronic devices of novel geometry could lead to potential applications for Van der Waals electronics.
\end{abstract}

\maketitle

When two electrically isolated conductors are brought close, a current in one conductor can generate friction and drag electrons in the other via Coulomb interaction, thereby causing a charge imbalance in the dragged layer. This is known as Coulomb drag and has been used to investigate the strongly correlated nature of low dimensional conductors \cite{narozhny_coulomb_2016}. Bilayers of two-dimensional (2D) electron gases
separated by only a few nanometers have been realized in semiconductor
hetero-structures \cite{gramila_mutual_1991} leading to several groundbreaking experiments such as the observation of a super-fluid exciton condensate \cite{eisenstein_boseeinstein_2004}: a quantum state carrying non-dissipative current of electron-hole pairs. More recently, electrical
devices based on hexagonal boron nitride (h-BN) encapsulated graphene \cite{gorbachev_strong_2012}
or bilayer-graphene \cite{li_negative_2016,lee_giant_2016} has allowed
the demonstration of Coulomb drag in 2D in previously inaccessible
regimes and  the formation of a quantum condensate at much higher temperatures
\cite{li_excitonic_2017,liu_quantum_2017}.  
In one-dimensional (1D) wires, Coulomb drag is also expected to reveal
a rich variety of exotic phenomena such as the formation of interlocked
charge-density wave \cite{fuchs_coulomb_2005} as well as Luttinger
liquid behavior \cite{flensberg_coulomb_1998}.

While in 2D the main observations of drag transport can be understood by assuming a weak Coulomb repulsion \cite{jauho_coulomb_1993}, this approach is believed to break down in 1D, where physics
of electrons is dominated by interaction. The limit of strong Coulomb
repulsion is reached in devices showing larger confinement such
as quasi-one-dimensional wires or in quantum dots \cite{keller_cotunneling_2016}. For instance, electrostatically defined nanowires have shown Coulomb drag signatures compatible with strongly interacting electrons such as negative drag \cite{yamamoto_negative_2006} and diverging behavior at low temperatures \cite{laroche_1d-1d_2014}. In this work, we present electron drag measurements in an asymmetric system composed of a carbon nanotube and graphene where electrons are propagating in 1D and 2D respectively. This system of hybrid dimensions allows the study of Coulomb drag in unexplored regimes beyond what has been observed in conventional systems  and brings interesting perspectives for the realization of a tunable and local drag probe for two-dimensional materials.  

\begin{figure}
\includegraphics[width=1\columnwidth]{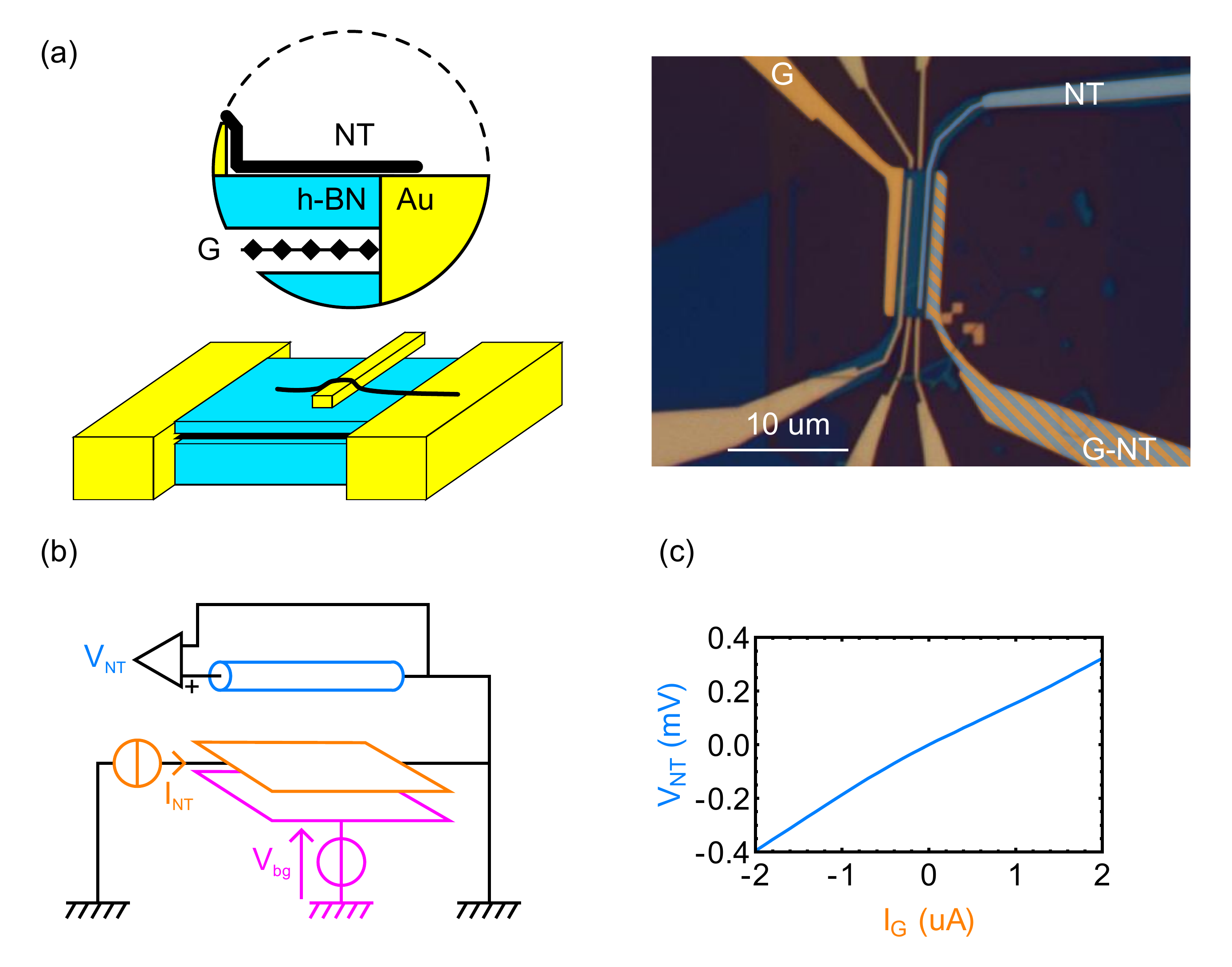}

\caption{\label{fig:Description-of-the-device}(a) Schematic of the carbon nanotube-graphene device and optical image of the device with false color on electrodes connected to graphene (orange) and carbon nanotube (blue). The device is measured in a cryostat equipped with a superconducting coil that can generate magnetic field up to 9~T. (b) Principle of Coulomb drag measurement across the nanotube. All measurements are performed using a 10 or 100 M$\Omega$ input impedance amplifier. (c) Typical $V_{NT}(I_{G})$ characteristic measured at temperature $T=1.6$~K with $V_{bg}=0$ applied on the back gate.}
\end{figure}

The device geometry used in this experiment is shown in Fig. \ref{fig:Description-of-the-device}a. A carbon nanotube is placed on top of an h-BN encapsulated graphene such that both conductors are electrically separated by a 12 nm thick h-BN. The device is made on an oxidized silicon wafer that we use as a back gate. Both nanotube and graphene are connected to metallic electrodes and have a common electrical ground in order to minimize uncontrolled voltage fluctuations.

\begin{figure}
\includegraphics[width=1\columnwidth]{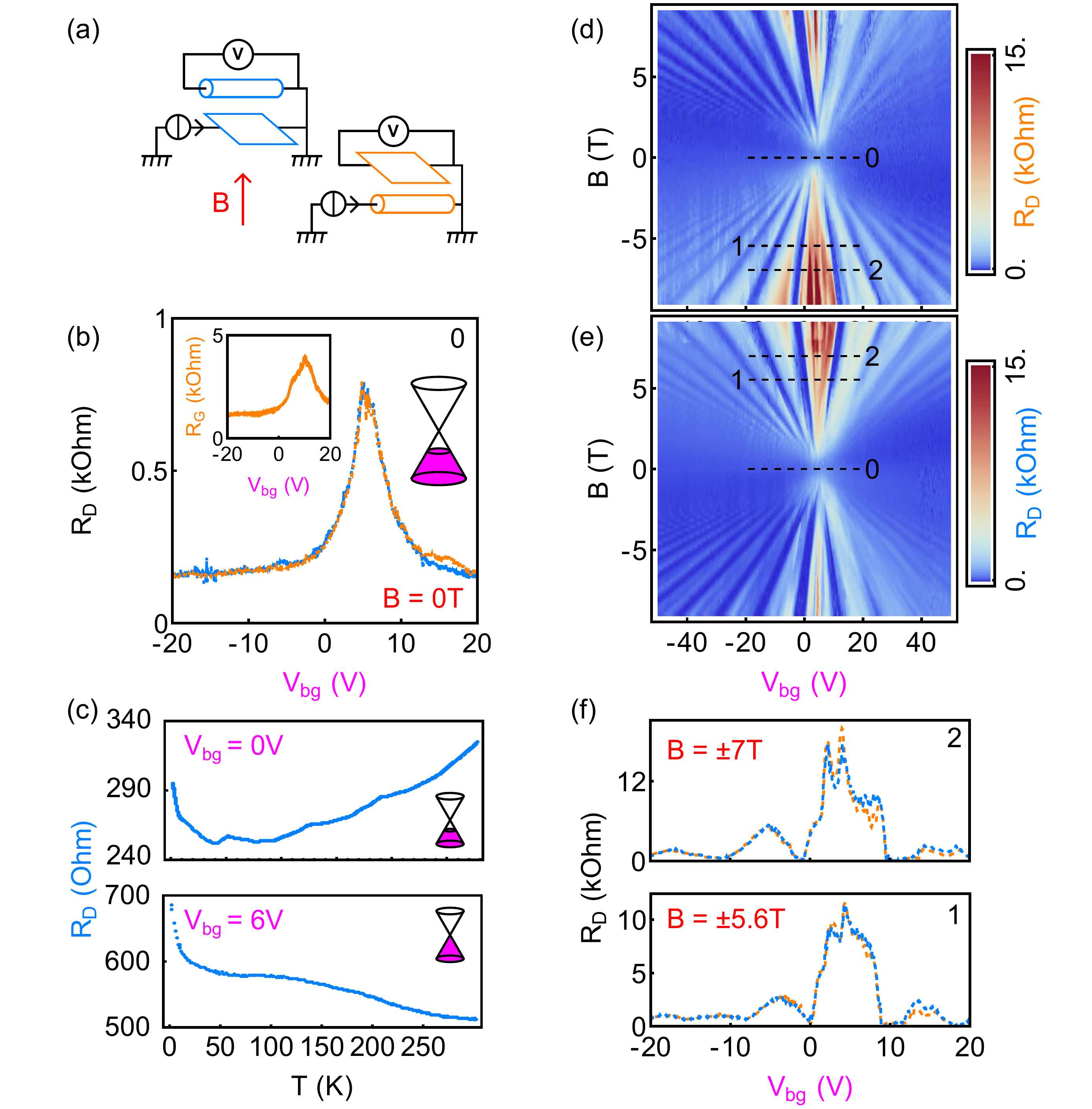}
\caption{\label{fig:Drag-response}(a) Schematic diagrams for two different configurations of drag measurements. The arrow indicates the direction of magnetic field $B$. (b) Measurements of $R_{D}$ as a function of back gate voltage $V_{bg}$ in absence of magnetic field. Left inset shows the resistance of graphene and right inset symbolizes the partially filled linear band structure of graphene. For the curve in blue, the drive current is flowing through the graphene and the voltage is detected across the nanotube. The orange curve shows similar drag measurement, the role of nanotube and graphene being inverted. (c) Evolution of $R_{D}$ with temperature $T$ for $V_{bg}=0$V (top, corresponding to hole doped graphene as shown in the inset) and 6~V (bottom, corresponding to the Dirac point as shown in the inset). (d)-(e) $R_{D}$ measured as a function of $V_{bg}$ and $B$ in the two configurations pictured in schematic diagrams in (a). We observe the formation of Landau levels when $B=\hbar\pi C_{SiO_{2}}(V_{bg}-V_{bg}^{0})/2e^{2}N$ where $\hbar$ is the reduced Planck constant, $e$ is the charge of electron, $N$ is an integer corresponding to the index of the Landau levels, $V_{bg}^{0}$ is the position of the Dirac peak and $C_{SiO_{2}}$ is the capacitance per surface unit between the back gate and graphene. The horizontal dashed lines show were cuts in other graphs are taken. (f) $R_D$ measured along the dashed lines shown in (d) and (e) at two different magnetic fields.}
\end{figure}

The principles of our experiment consist of flowing a current $I_{G}$
through graphene and detecting a voltage drop $V_{NT}$
across the carbon nanotube (Fig. \ref{fig:Description-of-the-device}b).
The slope of $V_{NT}(I_{G})$ characteristics (Fig. \ref{fig:Description-of-the-device}c) provides the drag resistance $R_{D}=dV_{NT}/dI_{G}$. This drag resistance can be modulated by the voltage applied to the back gate $V_{bg}$ reaching values of a few hundreds of Ohms. As illustrated in Fig. \ref{fig:Drag-response}a, the roles of graphene and carbon nanotube can be inverted, in which case the drag resistance is defined as $R_{D}=dV_{G}/dI_{NT}$ where $V_{G}$ is the voltage drop across the graphene and $I_{NT}$ is the current flowing through the nanotube. In both cases, because the nanotube is four orders of magnitude smaller than graphene and because the nanotube is mostly sensitive to the current flowing in its vicinity, drag measurements provide local information on the graphene properties.

Fig. \ref{fig:Drag-response}b shows $R_{D}$ as a function of $V_{bg}$ when there is no applied magnetic field $B$. $R_{D}$ shows a peak at $V_{bg}\approx 6$~V and  decreases slowly towards zero, following a power law (Supplementary Information (SI)),   for high positive and negative gate voltages. No appreciable difference can be found between two different configurations of measurement described above. At first approximation, we expect the back gate to barely affect the nanotube since the latter is screened by graphene. We therefore interpret the peak in $R_{D}(V_{bg})$ as a manifestation of the density of states minimum in the linear band structure of graphene, close to the Dirac point. It is informative to compare $R_{D}(V_{bg})$ with the gate dependent resistance of graphene $R_{G}(V_{bg})$ (inset of Fig. \ref{fig:Drag-response}b). While the charge neutral Dirac point of graphene can be roughly identified around $V_{bg}\sim 10$~V, close to the peak position of $R_{D}(V_{bg})$, we note that, however, $R_{G}(V_{bg})$ exhibits a broader, asymmetric peak with a poorly defined maximum  that we attribute to the charge inhomogeneity induced by the nanotube and local gating effect in the graphene channel coming from the nanotube contacts that are placed on top of h-BN   (SI) . In contrast, our drag measurements are local and therefore insensitive to such disorder. We also note that the sign of $R_{D}$ remains positive in the entire gate range.

The evolution of $R_D$ with temperature is a good indicator
to identify the relevant microscopic contributions. As seen in Fig. \ref{fig:Drag-response}c, away from the Dirac peak, the signal first decreases with temperature $T$ and then experiences a fast upturn with an amplitude change of approximately 25 percent, while at the Dirac peak, $R_{D}(T)$ increases continuously and seems to diverge as $T$ approaches 0. These observations are in a sharp contrast with the conventional drag in 2D conductors where drag resistance follows a $T^{2}$ law~\cite{gorbachev_strong_2012}. The conventional theoretical description of 2D drag is often based on a lowest order perturbation on the interaction strength \cite{narozhny_coulomb_2016}, and thus may not be applicable to our experiment, where strong Coulomb repulsion is expected due to confinement of electrons in the nanotube. Even though devices of hybrid dimensions such as ours have rarely been considered in literature \cite{lyo_coulomb_2003}, it is known that higher order terms calculated for 2D conductors yields non-vanishing drag contributions at low temperatures \cite{levchenko_coulomb_2008}, resulting in an increasing $R_D$ with decreasing temperature \cite{Schutt_coulomb_2013}. Beyond this perturbative approach, the Luttinger liquid theory also predicts a strongly non-monotonic behavior of $R_D$ (see comparison in supplementary) that largely depends on the microscopic details of the system \cite{flensberg_coulomb_1998,nazarov_current_1998,fuchs_coulomb_2005,fiete_coulomb_2006,dmitriev_coulomb_2012}.

\begin{figure*}
\includegraphics[width=0.75\textwidth]{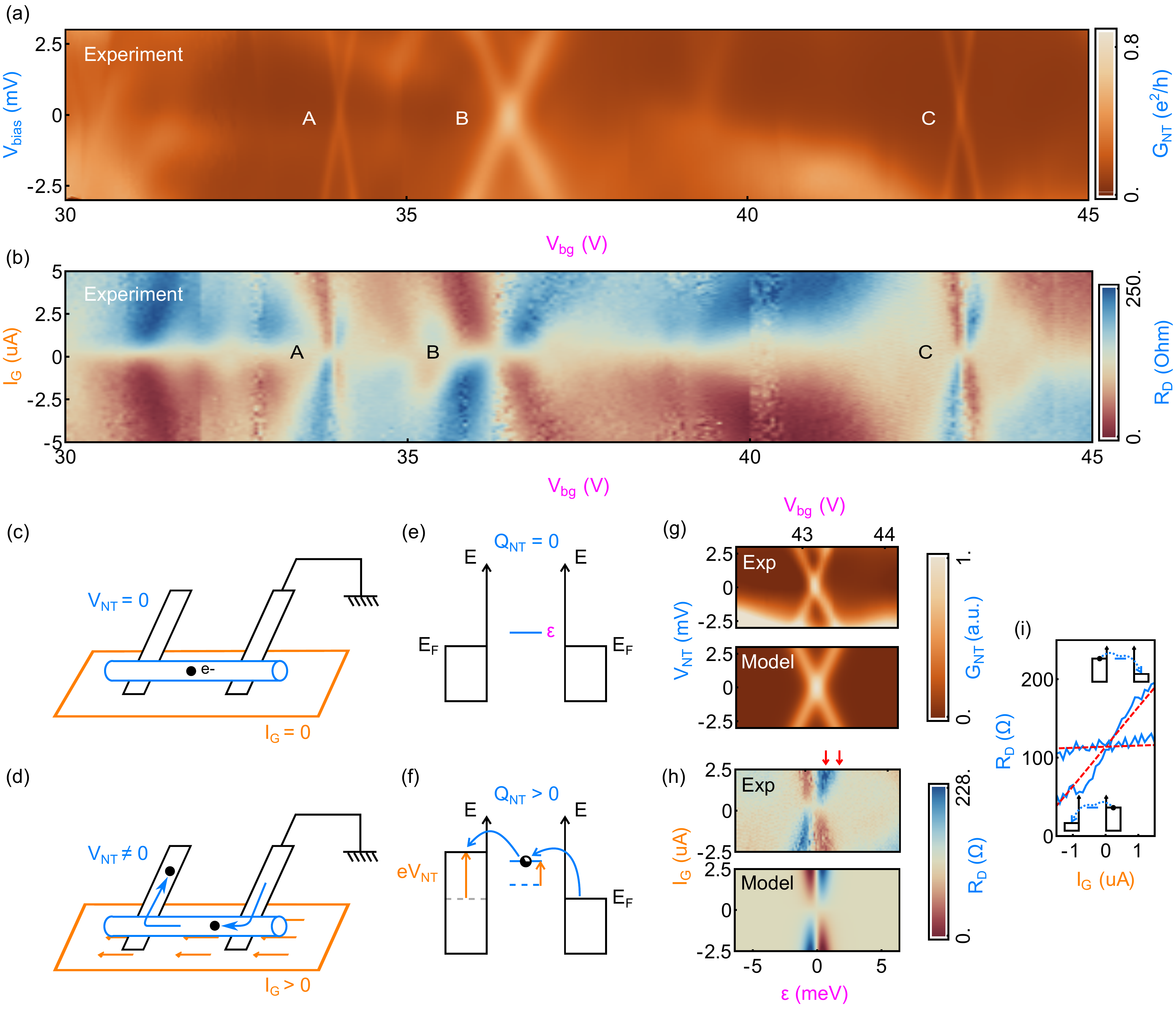}

\caption{\label{fig:Out-of-equilibrium-measurements}(a) Conductance of the nanotube $G_{NT}$ measured as a function of the voltage $V_{bias}$ applied across it in a range of back gate voltage $V_{bg}$ where graphene is highly doped and the effect of back gate
on the nanotube is weak. No current is flowing through graphene. (b) Drag resistance $R_{D}$ measured across the nanotube as a function of $V_{bg}$ and a finite DC current $I_{G}$ flowing through graphene. White color corresponds to median values of $R_{D}$ in the zero-bias limit, while blue and red respectively show increase and decrease of the drag resistance. (c) When no current is flowing through the graphene channel, charges are localized in the nanotube and no voltage develops across it. (d) If current is flowing, charges are displaced by an effective friction and $V_{NT}$ becomes finite. (e)-(f) If one considers a single empty electronic level of the nanotube with energy $\epsilon$, then a drag voltage developing across the nanotube changes the occupancy of this level. The energy cost associated to this process affects the magnitude of the drag signal. (g) Simulated conductance of the nanotube by solving master equations for $I_{G}=0$ as a function of the bias $V_{NT}$ and $\epsilon$. In our experiment $\epsilon$ can be controlled by $V_{bg}$. (h) Simulated drag resistance $R_{D}$ by solving the master equation for finite $I_{G}$ to obtain the derivative $dV_{NT}/dI_{G}$. The capacitance of the nanotube which control the amplitude of charge fluctuations is fitted at 16 aF. (i) Comparison between measured (blue) and simulated (red dashed) $R_D$ for two representative verical cuts marked by arrows in h. Near resonance, variations of $R_D$ are more significant.}
\end{figure*}

Upon finite magnetic field $B$ applied perpendicular to the graphene plane, we find that $R_D$ develops a rich structure as a function of $B$ and $V_{bg}$.  As shown in Fig. \ref{fig:Drag-response}d and e, the magnitude of $R_D$ increases with magnetic field as expected at low temperature \cite{Gornyi_coulomb_2004} (SI). On top of this increasing background level , $R_D$ also exhibits a series of oscillations that disperse with magnetic field and back gate voltage, fanning out in the $B$-$V_{bg}$ plane similarly to the Landau fan observed in $R_G$ (SI). In particular, we notice that $R_D$ is suppressed to nearly zero when the Fermi level of graphene is in between Landau levels. In this quantum Hall (QH) state, the electronic bulk of graphene becomes incompressible and currents are carried at the edges of the graphene channel. We find that $R_{D}$ decreases by up to three orders of magnitude as graphene enters the QH regime from more than 10~k$\Omega$ to only a few Ohms. This large modulation in $R_D$ is understandable if we consider the distance between the nanotube and the current: the nanotube is only 12 nm away from the current path when the bulk is conducting, whereas the distance becomes a few microns when the current is flowing along the edges.  These measurements also show that Onsager relations are not violated in our system as dictated by time-reversal symmetry (see Fig. \ref{fig:Drag-response}f and SI). Despite the physical asymmetry, $R_{D}$ is nearly identical when the role of nanotube and graphene as well as magnetic field are inverted.  

These measurements, taken in the linear regime and close to equilibrium, mainly reveal information on the graphene electronic states. However, away from this linear regime, we observe modulation of the drag resistance that is related to the nanotube internal electronic structure. This becomes apparent in the nanotube conductance measured as a function of the voltage bias applied across the nanotube (Fig. \ref{fig:Out-of-equilibrium-measurements}a).  At low $V_{bg}$ this structure is relatively smooth owing to the metallic nature of the nanotube, but it slowly evolves towards a Coulomb blockade regime at large gate voltage (SI). We then observe bright crosses (labeled A, B and C) corresponding to partial Coulomb diamonds revealing discrete electronic levels arising from spatial confinement as well as Coulomb repulsion. The smooth shape of these crosses suggests that the conductance $G_{NT}$ is dominated by a well-defined quantum dot but the uneven sizes of diamonds as well as the irregular background suggest a disordered environment. This structure also shows up in the drag resistance, when we drive the system out-of-equilibrium. Flowing a finite DC current in graphene $I_{G}$, we observe strong variations such that $R_{D}$ can increase by two fold or be completely suppressed (Fig. \ref{fig:Out-of-equilibrium-measurements}b). Comparing Fig. \ref{fig:Out-of-equilibrium-measurements}a and b, we identify that a suppression occurs when a nanotube electronic level is aligned with the Fermi level of the grounded electrode  whose potential is fixed, while enhancement occurs when it is aligned with the Fermi level of the floating one whose potential evolves freely.

The observed correlations between $R_D$ and $G_{NT}$ in the out-of-equilibrium regime can be described by a phenomenological model based on a friction force $\vec{F_{f}}=\eta n_{G}\vec{v}_{G}$ that graphene current exerts onto the localized electrons in the nanotube, where $\eta$ is the friction coefficient, $n_{G}$ is the graphene carrier density and $v_{G}$ is the velocity of electrons in graphene. We introduce an effective friction coefficient $\eta_{eff}$, such that this force can be written simply as $F_{f}=\eta_{eff}I_{G}$. When $I_{G}$ is finite, it displaces charges (see Fig \ref{fig:Out-of-equilibrium-measurements}c and d) across the nanotube from the grounded electrode to the floating one which consequently acquires a potential $V_{NT}$. During this process, each charge acquires an energy $-eV_{NT}+\eta_{eff}I_{G}L_{NT}$ where the first term comes from electrostatics and the second term is the work of friction along the length of the nanotube $L_{NT}$.   As friction increases the potential of the floating electrode $V_{NT}$, charge occupancy $\delta n_{NT}$ in the nanotube also increases (Fig. \ref{fig:Out-of-equilibrium-measurements}e and f). This charge increase comes into competition with the Coulomb repulsion within the nanotube. Consequently, as $\delta n_{NT}$ becomes larger (larger Coulomb energy), $V_{NT}$ is restrained, leading to a smaller $R_D$. Inversely, if occupancy decreases, an increase of $V_{NT}$ is favorable and $R_D$ becomes larger. These corrections are only significant when one electronic level is close to resonance with the Fermi level of the electrodes, as shown in Fig. \ref{fig:Out-of-equilibrium-measurements}i.  A more quantitative analysis can be done by employing a master equation formalism (see SI). Fig. \ref{fig:Out-of-equilibrium-measurements}g and h show the simulated $G_{NT}$ and $R_D$ respectively, across a single resonant level at energy $\epsilon$, which can be tuned by $V_{bg}$ in our experiment. Although we do not reproduce the smoother variations of the background, both simulated $G_{NT}$ and $R_D$ exhibit the essential features that appear in the data across a single resonant level.  We obtain a quantitative comparison with experiment using the nanotube capacitance as a single fit parameter, whose order of magnitude is imposed by the apparent sizes of Coulomb diamonds. Within this approach, we find an estimate for the friction coefficient $\eta$ of $1.8\hbar$ more than one order of magnitude larger than in graphene (SI) suggesting that, in this system, Coulomb drag is more prominent than in graphene-graphene devices.  

\begin{figure}
\includegraphics[width=1\columnwidth]{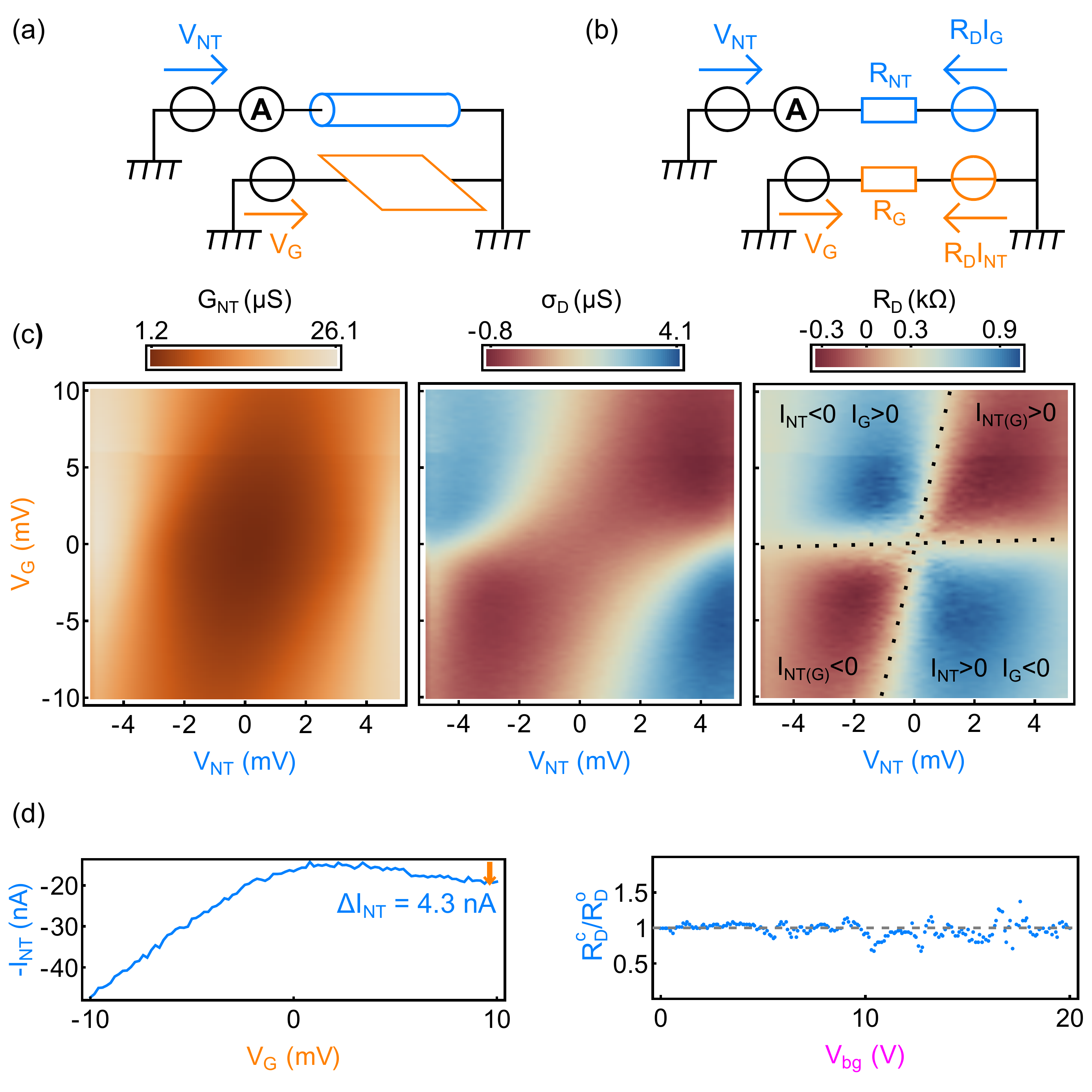}

\caption{\label{fig:Comparison-between-bias-and-drag}(a) Principle of simultaneous measurement of $G_{NT}$ and
$R_{D}$. Different frequencies are used for $V_{NT}$ and $V_{G}$ modulation to facilitate independent measurement. (b) Schematic used for the lumped element circuit analysis. (c) From left to right: differential conductance of the nanotube $G_{NT}$, drag conductance $\sigma_{D}$ and drag resistance $R_{D}=\sigma_{D}R_{NT}R_{G}$ where $R_{NT\left(G\right)}$ are the resistance of the nanotube and graphene, respectively. We can distinguish four quadrants in $R_{D}$ depending on the relative orientation of the current in graphene and nanotube. (d) On the left: DC measurements of $I_{NT}$ as a function of $V_{G}$ with $V_{NT}=3.2$ mV kept constant. The slope of the $I_{NT}\left(V_{G}\right)$ characteristic can change sign and the differential conductance $\sigma_{D}$ can thus be positive or negative. On the right: ratio of the differential drag resistance measured at $V_{G}=0$ across the nanotube in open circuit (Fig. \ref{fig:Drag-response}a) and in closed circuit (schematic a) with $V_{NT}=0$. These two different measurements yield consistent values.}

\end{figure}

The unconventional sign of $R_{D}$, its temperature dependence and the variations induced by charge fluctuations suggest an unconventional mechanism behind our observations that is possibly due to mesoscopic effects. Whether this mechanism is driven by charge repulsion \cite{moldoveanu_coulomb_2009,sanchez_mesoscopic_2010}, quantum shot noise \cite{levchenko_coulomb_2008-1} or non-local cotunneling processes \cite{kaasbjerg_correlated_2016}, the non-linearities in the nanotube conductance at a finite bias voltage can also produce non-linear drag behaviors. When both nanotube and graphene are biased separately as shown in Fig. \ref{fig:Comparison-between-bias-and-drag}a and b, the current in the nanotube $I_{NT}$ has two components: a standard resistive contribution proportional to the nanotube conductance $G_{NT} $ and a drag contribution proportional to $R_D$. Experimentally, we can distinguish these two contributions by frequency-division multiplexing with voltages having both a DC and an AC component. As shown in Fig. \ref{fig:Comparison-between-bias-and-drag}c at fixed gate voltage, the differential conductance of the nanotube $G_{NT}=dI_{NT}/dV_{NT}$ varies by more than one order of magnitude with $V_{NT}$, exhibiting a strong non-linear transport behavior but is weakly dependent on $V_{G}$. However, the drag conductance $\sigma_{D}=-dI_{NT}/dV_{G}$ varies rather strongly with respect to $V_{G}$ and $V_{NT}$. In this scheme, the non-linear drag resistance $R_D$ can be related to $\sigma_D$ through the relation: $\sigma_D \approx R_D/(R_{NT}R_G)$ (See Supplementary for derivation). The resulting drag resistance $R_{D}$ shows strong non-linear behavior including sign changes. In the right panel of Fig. \ref{fig:Comparison-between-bias-and-drag}c, $R_D$ shows four quadrants separated by boundaries along which the system is in the linear regime (i.e., in the limit of vanishing $I_{NT}$ and $I_{G}$) and $R_D$ takes constant value. Note that if $I_{G}$ and $I_{NT}$ have opposite directions, $R_{D}$ increases and stays positive but it decreases and reaches negative values for currents in the same direction. Such negative differential response, which can be explicitly demonstrated in DC measurement as well (Fig. \ref{fig:Comparison-between-bias-and-drag}d), could be of interest for the development of on-chip active devices   and detectors for different two-dimensional materials beyond graphene such as transition metal dichalcogenides or surface states of topological insulators.  

J.-D.P. would like to thanks \c C. Girit and C. Dean for fruitful discussions. This work is supported by ONR MURI N00014-16-1-2921 and Global Research Laboratory Program (2015K1A1A2033332) through the National Research Foundation of Korea (NRF) funded by the Ministry of science, ICT and Future Planning (MSIP). K.W. and T.T. acknowledge support from the Elemental Strategy Initiative
conducted by the MEXT, Japan and JSPS KAKENHI Grant Numbers
JP26248061,JP15K21722 and JP25106006.

\end{document}